\documentclass[floatfix,letterpaper,twoside,twocolumn]{revtex4}
\usepackage[utf8]{inputenc}
\usepackage{graphicx}

\newcommand{\Fig}[1]{Fig.~\ref{#1}}
\newcommand{\avg}[1]{<#1>}
\newcommand{\mrv}{{\bar\omega}}
\newcommand{\msrv}{\avg{\bar\omega^2}}

\newcommand{\amdoc}[1]{ Auxiliary Document #1}
\newcommand{\amvid}[1]{ Video #1}

\begin{document}

\title{Rotational transport via spontaneous symmetry breaking in
  vibrated disk packings}

% Use letters for affiliations, numbers to show equal authorship (if
% applicable) and to indicate the corresponding author
\author{Cristian Fernando Moukarzel}
\affiliation{ Departamento de Física Aplicada, CINVESTAV del IPN, 97310
  Mérida, Yucatán, México }
\author{Gonzalo Peraza-Mues} 
\affiliation{ Universidad Politécnica de Yucatán, Carretera
  Mérida-Tetiz. Km 4.5, Ucú, Yucatán, México }

\author{Osvaldo Carvente}
\affiliation{Departamento de Ingeniería Física, Universidad Autónoma de
  Yucatán, Mérida, Yucatán, México.}

\begin{abstract}
  It is shown that vibrated packings of frictional disks self-organize
  cooperatively onto a rotational-transport state where the long-time
  angular velocity $\bar\omega_i$ of each disk $i$ is nonzero.  Steady
  rotation is mediated by the spontaneous breaking of local reflection
  symmetry, arising when the cages in which disks are constrained by
  their neighbors acquire quenched disorder at large packing
  densities.  Experiments and numerical simulation of this unexpected
  phenomenon show excellent agreement with each other, revealing two
  rotational phases as a function of excitation intensity,
  respectively the low-drive (LDR) and the moderate-drive (MDR)
  regimes. In the LDR, interdisk contacts are persistent and rotation
  happens due to frictional sliding. In the MDR, disks bounce against
  each other, still forming a  solid phase. In the
  LDR, simple energy-dissipation arguments are provided, that support
  the observed dependence of the typical rotational velocity on
  excitation strength.
\end{abstract}

\maketitle
Modern civilization rests upon our ability to harness energy on large
scales. The size and complexity of human societies is strongly
correlated with their energy consumption rate. Population and
technological complexity developed, in relative terms, only slowly,
until practical thermal engines where
developed~\cite{TTTNO16}. Thermal engines allowed large amounts of
already available fuel energy to be used to produce work.  A thermal
engine transforms random thermal agitation into directed motion, and
is thus in essence nothing but a noise rectifier. Noise rectification
is clearly a process of fundamental importance, and not only for
energy generation. Much of our modern electronic technology, as well
as a large number of important biological processes, depend on
mechanisms that filter out noise, obtaining predictable behavior out
of randomness, down to the microscopic scale. This justifies the
continued interest in this field, and the search for novel noise
rectification processes~\cite{CHPFO93,KD-SURM99,RBMN02,RHITT14}

The second law of thermodynamics ensures that no device extracting
work from random motion can operate in thermal equilibrium. Transport
requires dissipation. The subtle way in which the second law
constrains the behavior of even the simplest noise rectifier has been
discussed in physics textbooks~\cite{FLTFL11}. In addition to the
breaking of time reflection-symmetry, implied by the existence of
dissipation, a spatial symmetry must as well be broken for
nonequilibrium transport to occur.  Most nonequilibrium systems that
are capable of unidirectional motion~\cite{MFTR93,BSSFI12,SPWDT92} or
rotation~\cite{TYACG05,EWERO10,APVAS13,ZPSOA16,SERRM18} have, by
design, a broken spatial symmetry. Such systems are said to have an
\emph{explicit} symmetry breaking.

Remarkably, however, some \emph{a priori} spatially symmetric
dissipative systems are as well able to sustain transport. Transport
is made possible in these systems by the \emph{spontaneous} breaking
of a spatial symmetry, that is, the fact that the system becomes
dynamically trapped within a non-symmetric subset of its
phase-space~\cite{RMUFI14}.  Transport mediated by a spontaneously
symmetry breaking (SSBT) is less common than that occurring on
explicitly non-symmetric systems, and often requires some degree of
fine tuning or cleverly targeted
design~\cite{JPCMM95,RKCBM99,MRSRE04}.
\begin{figure}[htpb]
  \centering
  \includegraphics[width=0.4950\linewidth]{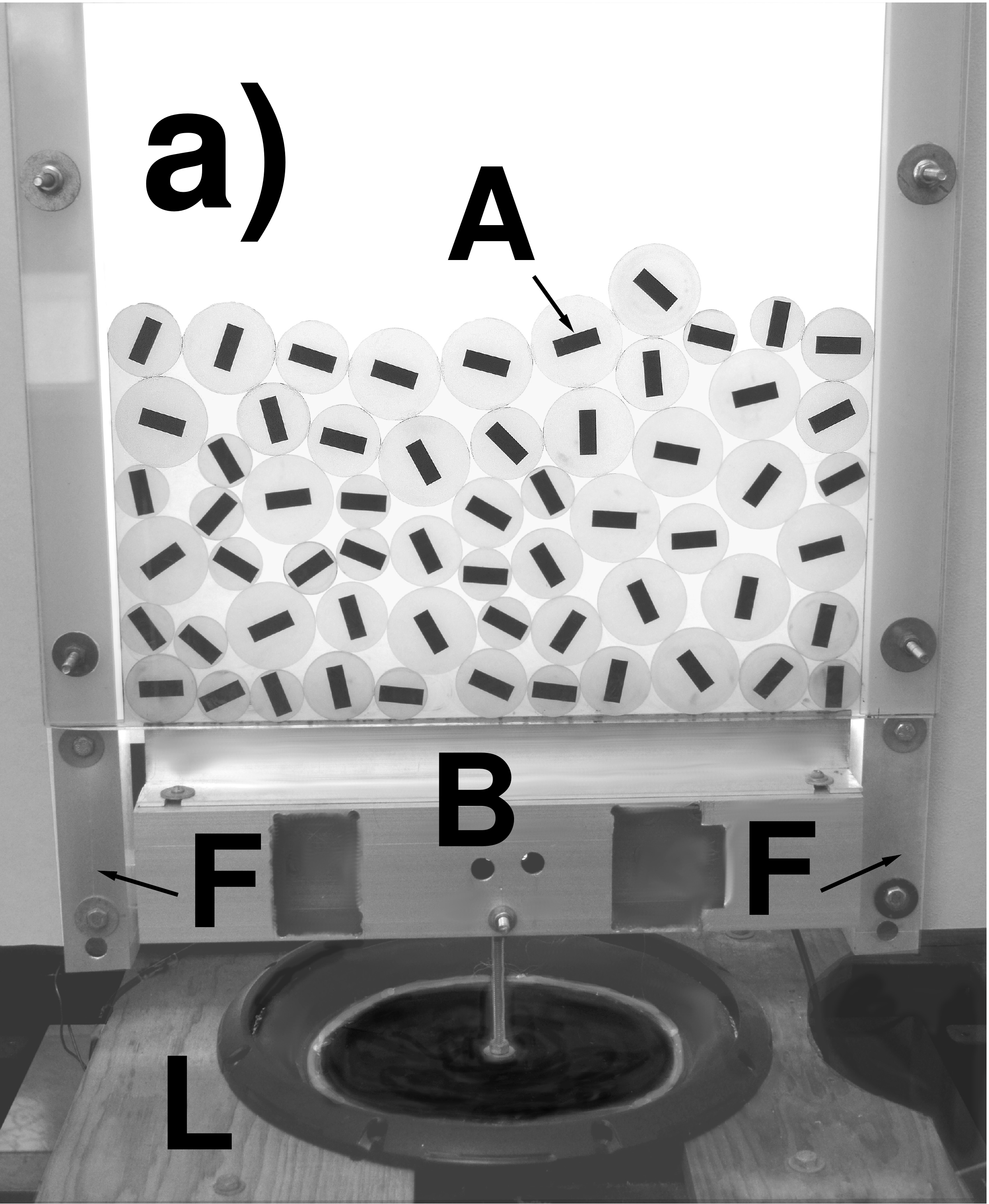}
  \begin{minipage}[b][0.22\textheight][s]{0.45\linewidth}
    \centering
    \includegraphics[width=0.995\linewidth]{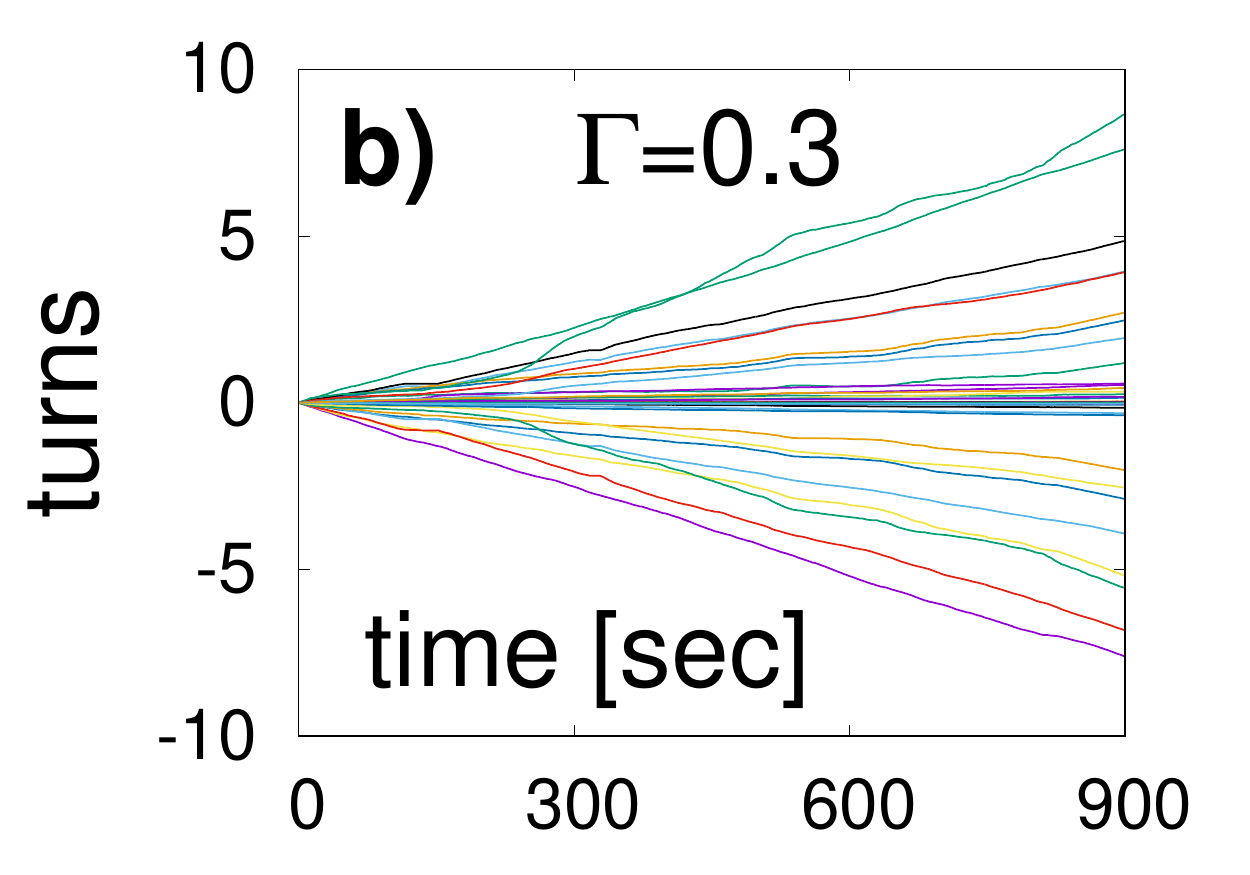}
    \vfill
    \includegraphics[width=0.995\linewidth]{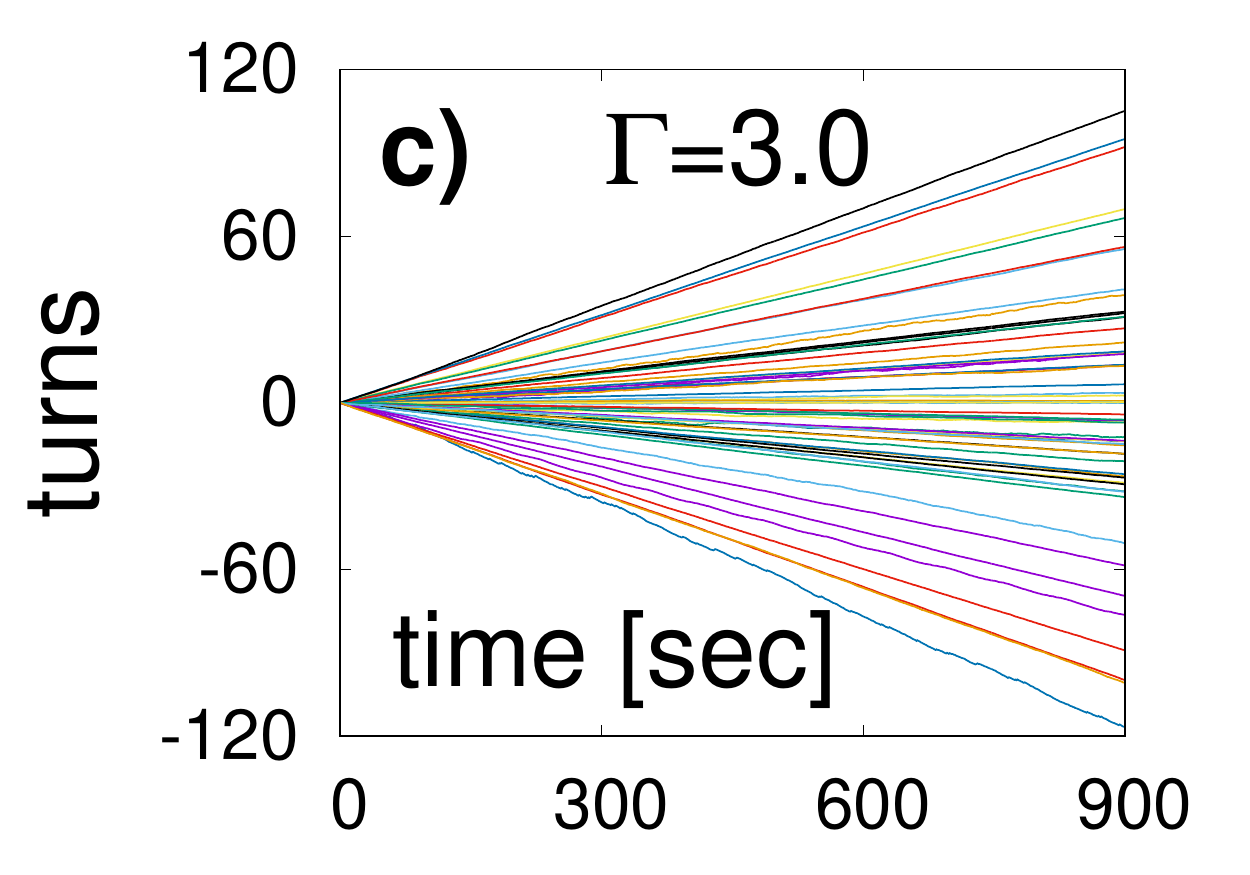}
  \end{minipage}
  \caption{\textbf{a)} Experimental apparatus used to study rotation
    in vibrated disk packings. A loudspeaker (L) rigidly connected to
    a movable bottom profile (B) provides the required
    excitation. Disks have labels (A) for motion tracking, and are
    confined between two glass panels rigidly fixed to the ground
    through metallic profiles (F). Experimental angular trajectories of
    all disks for a packing with (\textbf{b)} $\Gamma=0.3$,
    \textbf{c)} $\Gamma=3.0$) show the persistence of rotation over
    the observation timescale of 900 sec. }
  \label{fig:example-packing}
\end{figure}
Here, an unexpected novel example of SSBT is presented, that occurs in
a remarkably simple system: vertically vibrated disk packings under
gravity. We show, by experiment (See \amvid{Experiment}) and simulation
(See \amvid{Simulation}), that at large packing densities a
rotational-transport phase appears, in which each disk rotates with
nonzero long-time average angular velocity. Surprisingly, this
auto-organized transport phenomenon remained unnoticed despite decades
of intense work scrutinizing externally excited disks or spheres as
models of nonequilibrium phenomena, even when
rotations~\cite{MLENI98,HCADT05,BPTAR07,NDEOR12} where specifically
addressed in some cases.

Our main external control parameter is $\Gamma$, the peak bottom
acceleration divided by gravity $g$.  The disks, their mutual
interactions, and the external driving, are reflection-symmetric. At
large $\Gamma$, and therefore at low packing densities in our
gravity-confined setup (in the ``fluid'', highly agitated phase), no
rotational transport occurs, because the local environment of each
disk has chiral symmetry over long times. Systematic rotation requires
the dynamical breaking of this symmetry. This only happens, for the
system in question, at high packing densities. In the ``solid'' phase
that forms at high densities (small $\Gamma$), each disk becomes
nonergodically trapped in an irregular cage formed by its
neighbors~\cite{RICDI07}. Individual cages are not symmetric, and this
produces statistical imbalances in frictional interactions that make
systematic rotation possible over long times. The fact that most
earlier studies focused on the less dense fluid phase could explain
why rotation had not been reported previously.
\begin{figure}[htpb]
    \includegraphics[width=1.0\linewidth]{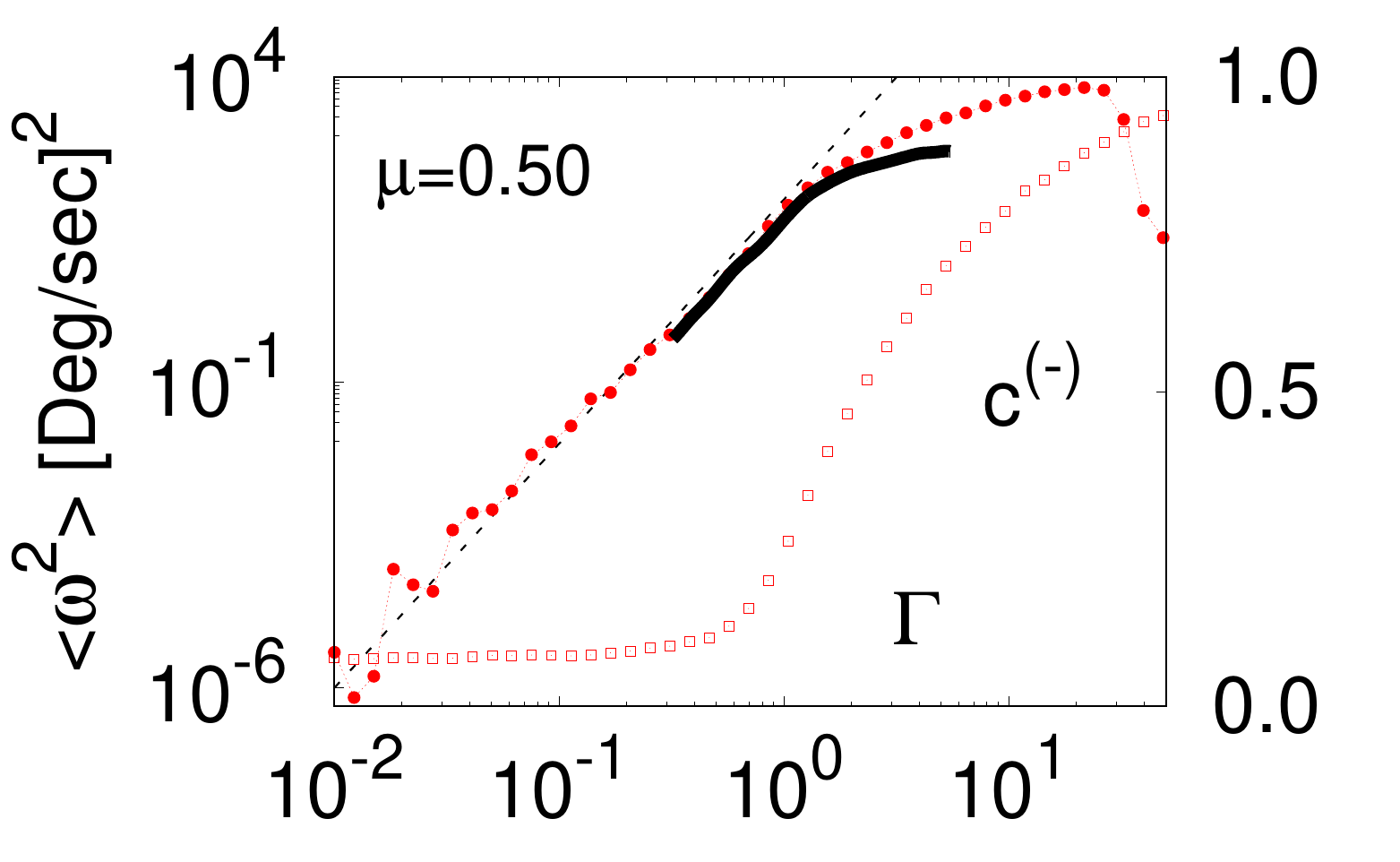} 
    \caption{ (color online) The mean squared rotational velocity
      \hbox{$\msrv$} (thick black line: Experiment, red full dots:
      Simulation) as a function of relative acceleration $\Gamma$
      reveals a dynamical transition at $\Gamma \approx 1$.  The
      nature of this transition is clarified by measuring the average
      contact-deficit $c^{(-)}$ defined in the text (red empty squares
      -- only in simulations). In the low-drive regime $\Gamma < 1$,
      contacts are persistent and disks rotate via sliding. The
      moderate-drive regime $1 < \Gamma \lesssim 30 $ is a
      collision-dominated rotational phase. The black dashed line is
      $\Gamma^4$. At very large drives $\Gamma > 30$, disk diffusion
      restores disk-neighborhood symmetry and $\msrv \to 0$ at long
      times. }
  \label{fig:1}
\end{figure}
\section*{Methods}
Experimental packings (\Fig{fig:example-packing}a) are made of 3 sets
of twenty disks each, respectively with radii $1.5, 2.0$ and $2.5$cm,
machine-cut from a 3mm-thick acrylic sheet, with sandpaper-treated
edges for increased friction, and an attached black rectangular label
(A) for motion tracking. Disks are confined between two vertical glass
panes (4mm separation), which are fixed to two steel frames (F) that
are rigidly attached to the floor.  Inserted through the 40-cm wide
bottom opening, an aluminum profile (B), restricted to move vertically
by a two-rail system, is rigidly connected to a loudspeaker (L) that
provides the vertical vibrational excitation. The speaker is driven by
a 1500W amplifier, fed by a sinusoidal signal with variable intensity
$S$ and fixed frequency $f_b=80$Hz.

After mixing the packing, the signal intensity is set to its maximum
admissible value $S=1$, and later decreases step-wise at 20-minute
intervals, until $S=0.05$ is reached. A set of $M$ measurements with
decreasing signal intensities $S_m$, $m=1,\ldots, M$ constitutes a run
$r$ of the experiment. Within each 20-minute interval, a 5-minute
relaxation period is followed by data acquisition for 15
minutes. Photos of the packing are taken at every second, and data
from two \emph{adxl345} triaxial accelerometric sensors $(L,R)$,
attached to opposite ends of the bottom profile, are stored at 200
Hz. Collision noise is filtered out from sensor signals by calculating
the integrals $I_m^{L,R}$ of their spectral densities in a 10Hz-wide
window around $f_b$. Our estimators
\hbox{$\Gamma_m^{L,R} = \sqrt{2I_m^{L,R}}(2 \pi f_b)^2/g$} for
vertical excitation result from calculating the peak acceleration
corresponding to a sinusoidal signal with spectral density integral
$I_m^{L,R}$ and dividing it by $g$. These two estimators are then
averaged to obtain $\Gamma_m$, verifying that their difference is
never larger than a few percents. We also check that horizontal and
off-plane vibrations are negligible.

Experimental data is presented from \textbf{25} runs $r=1,\ldots, 25$,
each taking approximately 10 hours to complete. For each run $r$ and
signal strength $S_m$, the resulting $T=900$ images are
processed~\cite{SAFAO12}, yielding positional and angular coordinates
\hbox{$\mathbf{x}^{r,m}_{i,t}, \theta^{r,m}_{i,t}$} for each disk $i$
at 1-sec intervals $t=1,\ldots, T$.

Although the same set $\{S_m\}$ of amplifier signals is always used in
experiments, the amplitude of motion of the bottom depends on the
response of each packing. Therefore, each run $r$ results in a
slightly different set of estimated vertical accelerations
$\{\Gamma^r_m\}$, where $m$ indicates the excitation amplitude.  To
estimate the $\Gamma$-dependence of measured quantities of interest,
averages are then taken over \emph{all} available measurements, using
a normalized Gaussian weight depending on \hbox{$(\Gamma^r_m-\Gamma)$},
with a narrow width $\sigma$.

In Molecular Dynamics (MD) simulation of disks, linear elasticity is
assumed, and the Cundall-Strack~\cite{CSADN79} algorithm is used to
calculate elastic tangential forces $\tau$. The algorithm assumes that
disks have an elastic ``skin'' that can be tangentially
stretched. Amonton's condition $|\tau| \leq \mu n$ constrains
tangential forces to be smaller than the normal force $n$ times the
friction coefficient $\mu$, which is appoximately 0.5 for
sandpaper-treated acrylic. A skin slip occurs whenever this condition
is violated, whereby some of the skin elastic energy is dissipated. A
fifth-order predictor-corrector algorithm~\cite{GTAI71} integrates
Newton's equations of motion in discrete time steps
$\delta t=10^{-6}$sec, which is approximately one hundredth of the
smallest elastic oscillation period in the system.  Physical
parameters such as normal elastic constant, mass density, etc.~where
chosen to represent the acrylic-disk experimental system as closely as
possible.  Further details of the simulation protocol follow standard
practices in MD simulations~\cite{FSUMS01,PSCGD05}.  A total of 224
different packings were simulated with MD with physical run times of
$t=5000$ sec, or $5 \times 10^{9}$ integration steps, per simulation
under constant conditions (the reasons for using such large simulation
times are explained in  \amdoc{2}  ).
The numerical results we
report here are averages over runs $r$, at constant $\Gamma$.
\section*{Results}
Figures \ref{fig:example-packing}b,c show that the angular coordinate
of each disk $i$ evolves almost linearly in time, with an average
angular velocity given by
\hbox{$\mrv_{i}= (\theta_{i,t=T}-\theta_{i,t=0})/T$}. Since
$\{\mrv_{i}\}$ are symmetrically distributed around zero, the typical
rotational velocity at a given level of excitation $\Gamma$ can be
estimated as the square root of $\msrv$, where $<>$ indicates an
average over all disks.

Our main results for $\msrv_{(\Gamma)}$, summarized in \Fig{fig:1},
show the existence of a dynamical transition near
\hbox{$\Gamma \approx 1$}, with very good phenomenological agreement
between experiment (thick line) and simulation (full dots). In both
cases $\msrv$ is found to behave as $\Gamma^4$ up to
$\Gamma \approx 1$, which is a natural driving
threshold~\cite{EWPDO07} under gravity. For larger excitation
intensities, $\msrv$ grows slower with $\Gamma$.  For very strong
intensities $\Gamma > 30$ (only in MD - See \amvid{High Drive} -- this
regime was not reachable in experiments), the packing undergoes
vibrofluidization, disks execute unbiased rotational random walks, and
$\msrv \to 0$ in the long run. We verified that the excitation
intensity at which this happens roughly coincides with the appearance
of diffusive behavior in disk centers.

Numerical simulation provides access to contact details, which allows
us to get useful insight into the nature of the dynamical transition
seen at \hbox{$\Gamma \approx 1$}.  We measure the \emph{Contact
  Deficit}, defined here as
\hbox{$ c^{(-)} =1-N_{contacts}/(2 N_{disks})$}, where $N_{contacts}$
is the number of disk contacts (walls included) at a given instant,
and $2N_{disks}$ is the isostatic~\cite{MIPT98} number of contacts
that defines a two-dimensional rigid packing in static
equilibrium. The Contact Deficit $c^{-}$ is zero for a static rigid
packing under gravity (in the limit of rigid, undeformable
disks~\cite{MIPT98}) , and equals one for a diluted granular gas with
no contacts. Its average value, obtained from MD simulations
(\Fig{fig:1}, empty squares) remains very small in the low-intensity
regime $\Gamma \lesssim 1$, growing steeply for $\Gamma \gtrsim
1$. This shows that in the low-drive regime the maximum possible
number $2N_{disks}$ of interdisk contacts remain closed most of the
time. The moderate-drive $\Gamma \gtrsim 1$ phase is, on the other
hand, a ``collisional'' phase where contacts occurr only briefly so
$N_{contacts} << 2N_{disks}$.

Armed with this knowledge, we can further our theoretical
understanding of the dynamical behavior at low excitation intensities.
In the low-drive regime $\Gamma \lesssim 1 $, an energy-dissipation
argument can be built, that supports the
\hbox{$\msrv \propto \Gamma^4$} relationship observed in \Fig{fig:1}
for experiments and simulation (See \amdoc{1}). The power dissipated by the packing
depends, among other parameters, on material properties such as
friction, elastic constants, and viscoelasticity, but we will only be
interestd in its $\Gamma$-dependence.  At constant driving frequency
$f_b$, the bottom, moving with amplitude $A_b$, injects an amount of
power $ P_{I} = C_{v} (f_bA_b)^2$ into the packing, with $C_{v}$ an
effective viscous coefficient depending on material properties of the
packing.  This power is dissipated, in general, through inelastic
collisions and frictional tangential forces $\tau$ among disks. In the
persistent-contact regime, however, dissipation due to inelastic
collisions can be neglected. The dominant dissipation mechanism is in
this case frictional, i.e.~due to contact slips. The global rate of
dissipation $P_{d}$ is thus roughly proportional to the average rate
of tangential slip times the average interdisk tangential force. The
average tangential force equals $\mu$, the frictional coefficient,
times the average normal force.  The globally averaged tangential slip
per unit time can be assumed to be proportional to the typical
rotational velocity, which is given by $\sqrt{\msrv}$.  The dissipated
power $\sim A_b^2$ is thus roughly proportional to
$\tau_{avg} \sqrt{\msrv}= \mu n_{avg} \sqrt{\msrv}$, where $n_{avg}$
is the time-averaged normal force between two disks, and averaged over
pairs of disks in contact.  Now, it is known that the hydrostatic
hypothesis is valid for excited granular systems~\cite{KCEFW14} such
as ours.  Therefore, interdisk normal forces are dictated by the
weight of the layers above a given pair, and thus do not vary with
$\Gamma$. Equating dissipated with injected power, using the fact that
$\Gamma \propto A_b$ and focusing on $\Gamma$-dependences only, one
obtains \hbox{$\msrv \propto A_b^4 \propto \Gamma^4$}, as observed in
the low-drive regime for both experiments and simulation (See
\Fig{fig:1}). This approach ceases to be useful in the bouncing regime
$\Gamma \gtrsim 1$, where free-flight rotation can no longer be linked
to dissipation.

By measuring the distance dependence of the two-disk angular velocity
correlation functions, we find that neighboring disks have a
statistical tendency to rotate in opposite directions, as favored by
frictional constraints. This of course does not imply perfect
slip-less rotation, which is only possible on arrays specifically
designed~\cite{SAPAC16} for that purpose.  As a further consistency
test for numerical results, Event-driven (ED)
simulations~\cite{PSCGD05} where also conducted for comparison with
MD. ED simulations only make sense at not too small driving
intensities ($\Gamma > 2$ in this case), so that persistent contacts
do not exist. We found a very good coincidence of results between
these two simulational methods, in the driving regime where both MD
and ED can be applied.

Given that our MD implementation is well validated by our own
experimental and ED simulation studies, we can use it as an
exploratory tool for more general setups that are not easily accesible
to experimentation. We did, for example, MD simulations of vibrated
elastic frictional packings with periodic boundary conditions across
the horizontal direction. The rotational properties discussed here
remain basically unchanged. We then conclude that rotation is not
wall-induced.  Other kinds of bottom excitation mechanisms were also
explored in simulations, as for example applying random uncorrelated
bumps on each disk of the lowest layer. These showed that rotational
organization does not depend on the periodic coordinated motion of the
bottom. The rotational phenomenon reported here was also found
numerically in closely confined packings without gravity, when these
are excited either by vibrating all four boundary walls or \emph{via}
Langevin noise (random point forces -- no torques) acting on disk
centers. This demonstrates that the rotational organization phenomenon
is cooperatively induced by frictional interactions among disks and
does not depend on the existence of a preferred gravity direction. We
plan to study horizontally confined 2D packings experimentally in the
near future.  More details on this rotational phenomenon will be
provided somewhere else for reasons of space.

We have made some analytical progress as well, for highly simplified
situations. We studied~\cite{PCRIA16,PMSRI19} three-disk toy
models~\cite{MGIAT05} or \emph{billiards}, consisting of one freely
moving disk supported against gravity by two vibrating ones. For
these, rotation invariably appears whenever the system is tilted with
respect to gravity, thus explicitely breaking reflection
symmetry. Rotation is seen in both the
persistent-contact~\cite{PMSRI19} (low drive) and
bouncing~\cite{PCRIA16} (moderate drive) regimes, which are also
observed for this simplified toy model.

In summary, we report a spontaneous-rotation phenomenon in vibrated
disk-packings under gravity.  Disorder in disk neighborhoods, in the
solid phase, locally breaks reflection symmetry, allowing for
persistent unidirectional rotation of disks. A similar rotational
transport phenomenon is observed in confined two-dimensional packings
without gravity. Based on our present understanding of this
auto-organized rotational state, we believe that a similar phenomenon
should be observed in sphere packings as well. This spontaneous
rotation phenomenon is potentially relevant for colloids, cellular
systems, and other close-packed systems of approximately circular or
spherical shape.

GPM was supported by a PhD fellowship from CONACYT México.  We are
grateful for the continued use of extensive computational resources on
clusters ``Xiuhcoatl'' and ``Kukulcán'' of CINVESTAV.  Ana María
Vidales and Aldo H.~Romero-Castro helped in the early numerical stages
of this work. Their contributions are gratefully acknowledged.
%\bibliographystyle{unsrt}
%\bibliography{extracted}
%\bibliography{extracted,RotatingDisks,Moukarzel,Ratchets,TTT}

\end{document}